\documentclass[12pt]{article}

\usepackage{amsmath,amssymb,cite,float,fullpage,graphicx,nicefrac,slashed,authblk,stix}
\usepackage{multirow}
\usepackage[T1]{fontenc}
\usepackage[table]{xcolor}

\usepackage[normalem]{ulem}

\title{Fermionic Electroweak Two-Loop Corrections to Drell-Yan and Related Processes}
\author{Ayres Freitas}
\author{E. Jackson Wallace}
\affil{\small Pittsburgh Particle-physics Astro-physics \& Cosmology Center
(PITT-PACC),\\ Department of Physics \& Astronomy, University of Pittsburgh, Pittsburgh, PA 15260, USA}
\date{} 

\begin{document}

\maketitle

\begin{abstract}
We perform a complete calculation of the next-to-next-to-leading order (NNLO) electroweak fermionic corrections to fermion-pair production processes, where ``fermionic'' refers to contributions with closed fermion loops. We do this via a semi-numerical technique that uses dispersion relations for the fermion sub-loop in two-loop box and vertex diagrams and dispersion relations and Feynman parameters for vertex diagrams with fermionic triangle sub-loops. UV and IR divergences are treated with suitable subtraction terms. We present numerical results for the cross-sections of $e^+e^-\to \mu^+\mu^-/u\bar{u}/d\bar{d}$ and differential distributions at representative center-of-mass energies. The NNLO corrections are found to modify the NLO cross-section on the order of 1\%.
\end{abstract}


\section{Introduction}

The discovery of the Higgs boson in 2012 by CMS and ATLAS~\cite{ATLAS:2012yve,CMS:2012qbp} has been a tremendous success of the Standard Model (SM) of particle physics, but since then no new particle discoveries have been made. As particle physics enters the high-luminosity LHC era, searches for beyond-the-Standard-Model (BSM) physics will continue to probe ever smaller deviations between experiment and SM prediction, requiring improved theoretical precision for the latter. One particular avenue to explore BSM physics is by examining deviations either in the Drell-Yan processes ($q\Bar{q}\to Z/\gamma \to \ell\Bar{\ell}$ at parton level) at HL-LHC~\cite{ATLAS:2019mfr} or fermion pair production at a future collider such as FCC-ee~\cite{FCC:2018evy}, CEPC~\cite{CEPCStudyGroup:2018ghi}, or ILC~\cite{ILC:2013jhg,Bambade:2019fyw}.

In order to extract these cross-sections, the theoretical precision should be near the experimental precision. The first-order (NLO) corrections to Drell-Yan have been known for some time~\cite{Altarelli:1978id,Altarelli:1979ub,Baur:2001ze, Dittmaier:2009cr}, while the next-to-next-to-leading order (NNLO) QCD corrections are also well-established~\cite{Hamberg:1990np,Harlander:2002wh,Anastasiou:2003yy,Melnikov:2006kv,Catani:2009sm,Bozzi:2010xn}, along with tools for predicting cross-sections at colliders~\cite{CarloniCalame:2007cd,Gavin:2010az,Barze:2013fru,Boughezal:2016wmq,Grazzini:2017mhc}. More recently, results have been published for mixed electroweak--QCD NNLO corrections~\cite{deFlorian:2018wcj,Delto:2019ewv,Cieri:2020ikq,Dittmaier:2014qza,Dittmaier:2015rxo,Bonciani:2020tvf,Dittmaier:2020vra,Armadillo:2022bgm,Armadillo:2024ncf,Buccioni:2022kgy,Bonciani:2021zzf} and for the N3LO QCD corrections~\cite{Duhr:2020seh,Duhr:2021vwj,Chen:2021vtu,Chen:2022cgv}, also including threshold resummations~\cite{Ahmed:2014cla,Catani:2014uta,Neumann:2022lft}. 

\begin{figure}
   	\begin{tabular}{@{}cccc@{}}   \includegraphics[width=0.2\textwidth]{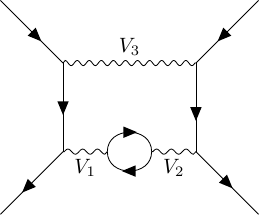} &  \includegraphics[width=0.23\textwidth]{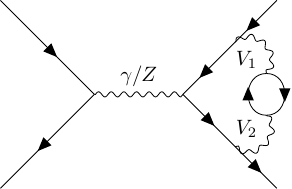} & \includegraphics[width=0.23\textwidth]{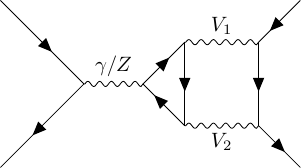} & \includegraphics[width=0.22\textwidth]{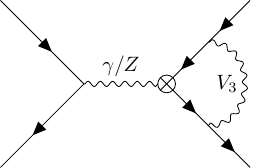} \\[-2ex]
    (a) & (b) & (c) & (d)
    \end{tabular}
  	\caption{Example diagrams for second-order electroweak corrections to fermion scattering. $V_i=\gamma/Z/W$ indicates a SM gauge boson. The cross in diagram (d) indicates a counterterm.}
	\label{fig:secondorderfermion}
\end{figure}

The next largest set of contributions to the Drell-Yan cross-section off the Z pole that have yet to be calculated are those from two-loop electroweak diagrams. Here one can further focus on contributions where at least one of the loops is a closed fermion sub-loop, which we expect to be relatively large due to the mass of the top quark and the number of light fermion degrees of freedom in these loops. The contributions near the Z pole can be treated using a Laurent expansion around the complex pole as described in~\cite{Chen:2022dow}. For energies sufficiently far above the Z pole, however, one needs to consider the full $ff \to f'f'$ process. Examples of diagrams with at least one fermion sub-loop are shown in Figure~\ref{fig:secondorderfermion}. Owing to the large number of mass scales in these diagrams, analytical methods based on a reduction to a set of master integrals can be challenging. However, purely numerical methods run into difficulties from the large dimensionality and complexity of the integrals.
    
To avoid these issues, in this paper, we use a semi-numerical method to reduce these calculations to up to two numerical integrals over Passarino-Veltman functions. In this way, the contribution of these diagrams to Drell-Yan scattering can be evaluated to high precision within minutes. In the following section, we provide more details of the calculational methods, the renormalization procedure, and the factorization of infra-red (IR) divergent contributions. In section \ref{sec:results}, we present numerical results for the partonic processes $ff \to f'f'$, where $f \neq f'$, and $f$ or $f'$ can be different combinations of lepton or quark flavors. We also provide an estimate of the uncertainty from missing electroweak corrections at NNLO without closed fermion loops and at N3LO.

\section{Description of the calculation}
\label{sec:calc}

The approach used in this paper is based on the techniques used in~\cite{Bauberger:1994by,Song:2021vru,Freitas:2022hyp}. Vertex and box diagrams with a sub-loop fermion self-energy, see Fig.~\ref{fig:secondorderfermion} (a,b), are evaluated using a dispersion relation for the self-energy. For most diagrams, only the transverse component of the self-energy will contribute to the cross-section. Thus, a two-loop diagram with a sub-loop self-energy takes the form~\cite{Du:2019evk}:
 \begin{align}
    \int\frac{d^Dk}{i\pi^{d/2}}\;\frac{N(k)}{\prod_i[(k+p_i)^2-m_i^2+i\epsilon]}\hat\Sigma^{V_1,V_2}_{\mu\nu}(k^2)    
\end{align}
where $V_{1,2}=W,Z,A$ denote different types of electroweak gauge bosons and $\hat\Sigma$ is the sub-loop self-energy with suitably subtracted sub-loop counterterms to make it finite:
\begin{align}
    \hat\Sigma^{AA/AZ}_{\mu\nu}(k^2) & = (g_{\mu\nu}k^2-k_\mu k_\nu)\hat\Pi^{AA/AZ}_T(k^2), \notag \\
    \hat\Sigma^{VV}_{\mu\nu}(k^2) & = \frac{1}{k^2}\left(g_{\mu\nu}k^2-k^\mu k^\nu\right)\hat\Sigma^{VV}_T(k^2), \notag \displaybreak[0] \\ 
    \hat\Pi^{AA}_T(k^2) & = \Pi^{AA}_T(k^2) + \delta Z^{AA} = \frac{k^2}{\pi}\int_0^\infty  d\sigma\; \frac{\text{Im}\{\Pi^{AA}_T(\sigma)\}}{\sigma(\sigma-k^2-i\epsilon)}\,, \label{eq:selfdisp} \\
    \hat\Pi^{AZ}_T(k^2) & = \Pi^{AZ}_T(k^2) + \tfrac{1}{2}\delta Z^{AZ} = \frac{k^2-m_Z^2}{\pi}\intbar_0^\infty d\sigma\; \frac{\text{Im}\{\Pi^{AZ}_T(\sigma)\}}{(\sigma-m_Z^2)(\sigma-k^2-i\epsilon)}\,, \notag \displaybreak[0] \\
    \hat\Sigma^{VV}_T(k^2) & = \Sigma^{VV}_T(k^2) -\delta m_V^2 + \delta Z^{VV}\,(k^2-m_V^2) = \frac{(k^2-m_V^2)^2}{\pi}\intbar_0^\infty d\sigma\; \frac{\text{Im}\{\Sigma^{VV}_T(\sigma)\}}{(\sigma-m_V^2)^2(\sigma-k^2-i\epsilon)}\,, \notag
\end{align}
where $\intbar$ indicates that the poles from $1/(\sigma-m_V^2)$ terms are to be evaluated as principal-value integrals, and $\delta m^2_V$ and $\delta Z^{V_1V_2}$ denote mass and field renormalization counterterms (see below for more information on the renormalization scheme).
For photon-photon and photon-Z self-energies, the contribution of a fermion of mass $m_f$ is given by
\begin{align}
    \text{Im}\{\Pi^{AA}_T(\sigma)\} = \text{Im}\{\Pi^{AZ}_T(\sigma)\} & = \frac{N_c g_1g_2}{12\pi}\biggl(1+\frac{2m_f^2}{\sigma}\biggr)\sqrt{1-\frac{4m_f^2}{\sigma}} \; \Theta(\sigma-4m_f^2).
\end{align}
Here $\Theta(x)$ is the Heaviside step function, $g_1g_2 = e^2Q_f^2$ (photon-photon) or $\frac{e^2Q_f(2s_W^2Q_f-I_{3f})}{2s_Wc_W}$ (photon-Z), and $N_c=1\,(3)$ for leptons (quarks).
For the sake of brevity, we do not write the corresponding expressions for the massive gauge boson self-energies. The masses of light fermions can be safely neglected in the AZ/ZZ/WW self-energies, but they need to be kept in the AA self-energy to regulate IR singularities.
    
For diagrams with triangle fermion sub-loops, it is convenient to use Feynman parameters to transform them into self-energy sub-loops.
\begin{align}
    I^{\rm v} &= \int_{-\infty}^\infty d^D q_2 \; \frac{q_2^\mu \cdots q_2^\nu}{\bigl[(q_1+q_2)^2-m_f^2\bigr]\bigl[q_2^2-m_f^2\bigr]\bigl[(q_2-p_1-p_2)^2-m_f^2\bigr]} \notag \\
    &= \int_0^1 dx \int_{-\infty}^\infty d^D q_2 \; \frac{q_2^\mu \cdots q_2^\nu}{\bigl[(q_1+q_2)^2-m_f^2\bigr]\bigl[(q_2-p')^2-{m'}^2\bigr]^2} \notag \\
    &= \int_0^1 dx \; \frac{\partial}{\partial (m')^2} \int_{-\infty}^\infty d^D q_2 \; \frac{q_2^\mu \cdots q_2^\nu}{\bigl[(q_1+q_2)^2-m_f^2\bigr]\bigl[(q_2-p')^2-{m'}^2\bigr]} \notag \\
    &= \int_0^1 dx \; \Bigl[ \sum_k N_k^{\mu \cdots \nu}\,B_k(q_1^2,m_f^2,m'^2) \Bigr], 
\end{align}
with the integral now expressible in terms of simpler functions via typical Passarino-Veltman reduction. Here $q_1$ is the loop momentum of the second loop, $p_{1,2}$ denote external momenta, and
\begin{align}
    p' &=  (p_1+p_2)(1-x),  & m'^2 &= m_f^2-(p_1+p_2)^2x(1-x). 
\end{align}
Furthermore, $N_k^{\mu \cdots \nu}$ is some combination of metric tensors and $q_1$ four-vectors.

The $B_k$ functions are then also expressed via dispersion relations~\cite{Song:2021vru}
\begin{align}
    B_k(q_1^2,m_f^2,m'^2) &= \frac{1}{\pi} \int_0^\infty d\sigma \; \frac{\text{Im}\bigl\{ B_k(\sigma,m_f^2,m'^2) \bigr\}}{\sigma-q_1^2-i\epsilon} && (\text{for } m'^2>0), \\
    B_k(q_1^2,m_f^2,m'^2) &= \frac{1}{2\pi i} \int_{-\infty}^\infty d\sigma \; \frac{B_k(\sigma,m_f^2,m'^2)}{\sigma-q_1^2-i\epsilon} && (\text{for any } m'^2).
\end{align}

\begin{figure}
\centering
   	\begin{tabular}{@{}cc@{}}   \includegraphics[width=0.26\textwidth]{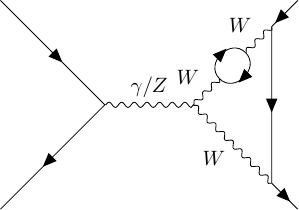} &  \includegraphics[width=0.26\textwidth]{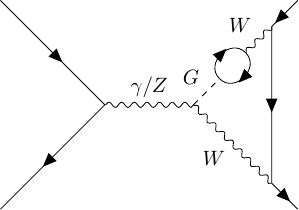} \\[-2ex]
    (a) & (b)
    \end{tabular}
  	\caption{Example diagrams of sub-loops with longitudinal contributions.}
	\label{fig:longitudinalvertex}
\end{figure}

For a specific set of diagrams where the vertex loop is composed of three W bosons as in Fig.~\ref{fig:longitudinalvertex}~(a), the longitudinal component of the self-energy will also contribute to the cross-section. These contributions are suppressed for diagrams where the vector bosons couple to external fermion lines since, due to the Goldstone equivalence theorem, they are proportional to the mass of this fermion. However, for Fig.~\ref{fig:longitudinalvertex}~(a) there is a W boson that does not couple to any massless fermions, and thus the longitudinal mode contributes. The sub-loop divergences cancel with the corresponding diagram with a Goldstone boson, as in Fig.~\ref{fig:longitudinalvertex}~(b).

The two-loop vertex diagrams can contain UV divergences, which must be handled prior to numerical integration. For the fermion sub-loop divergence, this is achieved by subtracting a term where all other momenta are set to zero in the $q_2$ integral~\cite{Freitas:2022hyp}:
\begin{align}
  I^{\rm v} &\to I^{\rm v} - I^{\rm v}_{\rm sub,2}, \label{eq:sub2a} \\
  I^{\rm v}_{\rm sub,2} &= \int_0^1 dx \; \frac{\partial}{\partial (\mu^2)} \int_{-\infty}^\infty d^D q_2 \frac{q_2^\mu \cdots q_2^\nu}{\bigl[q_2^2-m_f^2\bigr]\bigl[q_2^2-\mu^2\bigr]}, \label{eq:sub2b}
\end{align}
where $\mu^2$ is an arbitrary constant mass parameter. As a result, eq.~\eqref{eq:sub2a} is UV-finite and thus suitable for numerical evaluation in $D{=}4$. On the other hand, the integral eq.~\eqref{eq:sub2b} can be straightforwardly evaluated analytically in terms of well-known one-loop functions and added back to the whole expression. 

Similarly, there can be nested UV divergencies of the full two-loop integrals, which can be compensated by subtracting corresponding two-loop integrals with all external momenta set to zero. The latter can also be computed analytically in dimensional regularization in terms of known two-loop vacuum functions and added back.

\paragraph{Treatment of \boldmath $\gamma_5$:}
    
    Traces such as $\text{Tr}(\gamma^\mu\gamma^\nu\gamma^\rho\gamma^\sigma\gamma_5) = 4i\epsilon^{\mu\nu\sigma\rho}$ appear in the diagrams with a triangular fermion sub-loop. In general, the treatment of traces containing $\gamma_5$ in $D$ dimensions is ambiguous because the previous equation is incompatible with $\{\gamma_5,\gamma_\mu\}=0$ in $D\neq 4$.  However, terms containing an $\epsilon$ tensor are required to be UV-finite at two-loop order. As a result, we can carry out this calculation by splitting such diagrams into two parts: The first part is evaluated using a naively anti-commuting scheme in $D$ dimensions and $\text{Tr}(\gamma^\mu\gamma^\nu\gamma^\rho\gamma^\sigma\gamma_5) = 0$. Then, separately, the terms containing epsilon tensors can be evaluated in 4 dimensions, since they are UV-finite.

\paragraph{Renormalization procedure:} 
    
    The on-shell renormalization scheme is used in this work, using fermion masses, W/Z/H boson masses, and the electromagnetic coupling as independent parameters. Concretely, the on-shell masses are defined via the location of the pole of their radiatively corrected propagators. For particles with non-negligible decay widths, the propagator pole is complex, which affects the mass counterterms at NNLO and beyond. In our case, this needs to be considered for the W and Z boson masses, $m_{W,Z}$. Note that $m_{W,Z}$ defined in this way differ numerically from the mass definitions used in experimental analyses. The relation between them is given by
    \begin{align}
        \begin{aligned}
        m_Z &= m_Z^{\rm exp}[1+(\Gamma_Z^{\rm exp}/m_Z^{\rm exp})^2]^{-1/2} 
         \approx m_Z^{\rm exp} - 34\,\text{MeV}, \\
        m_W &= m_W^{\rm exp}[1+(\Gamma_W^{\rm exp}/m_W^{\rm exp})^2]^{-1/2} 
         \approx m_W^{\rm exp} - 27\,\text{MeV}.
        \end{aligned} \label{mztrans}
    \end{align}
    See Ref.~\cite{Freitas:2002ja,Freitas:2020kcn} for more details.
    
    For the renormalization of the electromagnetic coupling, three options are considered:
    \begin{itemize}
    \item \emph{$\alpha(0)$ scheme:} The electromagnetic coupling is defined via the $\gamma f\bar{f}$ vertex in the Thomson (zero momentum) limit. With this definition, predictions for electroweak processes depend on $\Delta\alpha \equiv 1-\alpha(0)/\alpha(m_Z)$, where $\alpha(\mu)$ is the running electromagnetic coupling at the scale $\mu$. $\Delta\alpha$ receives non-perturbative contributions from hadronic physics at (sub-)GeV scales and needs to be determined from data or lattice studies~\cite{pdg}.
    \item \emph{$\alpha(m_Z)$ scheme:} The running coupling at the $Z$ mass scale is used as input, which effectively resums the running effects from $\Delta\alpha$.
    \item \emph{$G_\mu$ scheme:} The electromagnetic coupling is derived from the Fermi constant $G_\mu$, using the relation $G_\mu = \frac{\pi \alpha}{\sqrt{2}(1-m_W^2/m_Z^2)m_W^2}(1+\Delta r)$. Here $\Delta r$ captures radiative corrections, which are known at full NNLO and partial higher orders~\cite{Awramik:2003rn}. For the purposes of this work, the NLO and fermionic NNLO corrections from Ref.~\cite{Freitas:2002ja} are utilized.
    \end{itemize}
    
\paragraph{Factorization of IR-divergent contributions:} 

    In general, diagrams such as those in Fig.~\ref{fig:secondorderfermion} (a,b,d) for $V_{1,2}=A$ or $V_3=A$ contain IR divergences originating from the phase-space region when internal photons become soft\footnote{Additionally, there are collinear divergences due to the fact that we set the external fermion masses to zero, but the pure collinear singularities cancel in the sum of all diagrams.}. These IR singular contributions have a universal structure, described by pure QED~\cite{Yennie:1961ad}, and factorize from the hard (non-QED) matrix element. They can be evaluated, together with the real radiation contributions, using Monte-Carlo (MC) techniques suitable for experimental simulations~\cite{Jadach:1998jb,Jadach:2013aha,Krauss:2022ajk}. Therefore, in this work, we focus on the hard NNLO matrix element that can be matched to such a QED MC tool. Accordingly, the universal IR-singular QED contributions must be subtracted from the hard matrix element to avoid double counting. [See Ref.~\cite{Armadillo:2025mfx} for more details on the structure of QED singularities and an alternative subtraction scheme from the one used here.]
    
    For initial- and final-state radiation, this can be achieved by simply dropping diagrams of the type in Fig.~\ref{fig:secondorderfermion} (d) with $V_3=A$. Such diagrams factorize into a QED radiator function $R(s)$, which would be captured by the MC simulation, and a hard vertex form factor, which is part of the NLO matrix element. Similarly, diagrams as in Fig~\ref{fig:secondorderfermion} (center) with $V_{1,2}=A$ are part of the universal QED corrections and thus should be excluded from the hard matrix element.
    
    Now let us consider box diagrams to the process $f\bar{f} \to f'\bar{f}'$ of the form in Fig~\ref{fig:secondorderfermion} (a) with $V_3=A$. Their universal IR-singular contribution is split off by the following subtraction:
    \begin{align}
        {\cal M}_{\rm box(2)}^{V_1=A}(s,t) \to {\cal M}_{\rm box}^{V_1=A}(s,t)
        - \frac{2\alpha}{\pi} Q_f Q_{f'} \big[ B_{(1)}(t) - B_{(1)}(u) \bigr] {\cal M}_{\rm se(1)}(s), 
    \end{align}
    where ${\cal M}_{\rm se1}(s)$ is the contribution to $f\bar{f} \to f'\bar{f}'$ from a one-loop diagram with $V_{2,3}$ and a fermion self-energy in the s-channel, and $B_{(1)}(Q^2)$ is the universal eikonal one-loop vertex factor \cite{Yennie:1961ad}
    \begin{align}
        B_{(1)}(Q^2) = \int \frac{d^Dq}{i(2\pi)^{D-2}}\; \frac{(q+2p_1)\cdot(q-2p_2)}{[q^2+i\epsilon]
        [(q+p_1)^2-m_f^2+i\epsilon][(q-p_2)^2 -m_f^2+i\epsilon]} 
    \end{align}
    with $p_1^2=p_2^2=m_f^2$ and $(p_1+p_2)^2=Q^2$.

    Similarly, the IR matching of diagrams of the form in Fig~\ref{fig:secondorderfermion} (left) with $V_{1,2}=A$ is carried out by the subtraction
    \begin{align}
        {\cal M}_{\rm box(2)}^{V_{2,3}=A}(s,t) \to {\cal M}_{\rm box(2)}^{V_{2,3}=A}(s,t)
         - \frac{2\alpha}{\pi} Q_f Q_{f'} \big[ B_{(2)}(t) - B_{(2)}(u) \bigr] {\cal M}_{\rm tree}(s), \label{eq:IR2b}
    \end{align}
    where ${\cal M}_{\rm tree}(s)$ is the tree-level matrix element with $V_1$ in the s-channel, and $B_{(2)}(Q^2)$ is the two-loop QED vertex factor
    \begin{align} 
        B_{(2)}(Q^2) = \int \frac{d^Dq}{i(2\pi)^{D-2}}\; \frac{(q+2p_1)\cdot(q-2p_2)}{[q^2+i\epsilon]
         [(q+p_1)^2-m_f^2+i\epsilon][(q-p_2)^2 -m_f^2+i\epsilon]} \, \Pi^{AA}_T(q^2).
    \end{align}
    For technical purposes, we introduce a fictitious photon mass $m_\gamma$ to regulate the IR divergences in individual diagrams, and we have verified that our results are independent of the value of $m_\gamma$ after performing the IR matching as described above. 
    
    Two-loop box diagrams as in Fig~\ref{fig:secondorderfermion} (left) with $V_{1,2}=A$ are also sensitive to the masses of light fermions in the self-energy loop, and it is not possible to take the limit $m_{lf} \to 0$ ($lf = e,\mu,\tau,u,d,s,c,b$). Even after performing the subtraction in eq.~\eqref{eq:IR2b} there is a residual dependence for small values of $m_{lf}$. This dependence is expected to cancel when adding the contribution of tree-level diagrams with an extra light fermion pair in the final state. However, for the virtual corrections by themselves, one must assign non-zero masses to the light fermions inside of photon self-energy sub-loops. For light quarks, their masses as elementary particles are not well-defined, and we illustrate the dependence on the chosen values in the numerical results. Even so, we reiterate that this dependence on the light quark masses should be reduced to a negligible level when adding contributions with a light quark pair in the final state, assuming that the same quark regulator masses are used there.

\paragraph{Implementation of calculation:}
The calculation has been carried out with two independent implementations for the different building blocks so as to allow cross-checking between us. We used \textsc{FeynArts 3.11}~\cite{Hahn:2000kx} to generate amplitudes based on the topologies of interest to this calculation. Dirac and Lorentz algebra were handled by \textsc{FeynCalc 9.3}~\cite{Shtabovenko:2020gxv} in one of our implementations and with a private code in the other. Both implementations used separate private \textsc{Mathematica} codes for expressing the amplitudes in terms of dispersion relations, constructing analytically integrable subtraction terms to handle UV divergences, and for the treatment of universal eikonal QED contributions. The numerical evaluation of the dispersion relation integrals was performed in \textsc{C++}, with one using the GNU Scientific Library~\cite{Galassi:2009} and the other using the \textsc{QuadPack}~\cite{quadpack} library. For the numerical integration, as well as for the terms we compute analytically and add back to the expression, the UV-finite Passarino-Veltman functions are evaluated using \textsc{LoopTools 2.16}~\cite{Hahn:1998yk}. Additionally, we used \textsc{TVID}~\cite{Bauberger:2019heh} for the evaluation of two-loop vacuum functions.


\section{Numerical results}
\label{sec:results}

This section presents results for electroweak corrections to partonic cross-sections. IR-divergent QED contributions are excluded according to the factorization scheme described in the previous section. For definiteness, we consider an $e^+e^-$ initial-state, with different final state lepton or quark pairs ($\mu^+\mu^-$, $u\bar{u}$, $d\bar{d}$).

These results are also applicable to Drell-Yan production at hadron colliders, by simply swapping initial and final states for $e^+e^- \to u\bar{u}/d\bar{d}$. However, the full phenomenological description of either process would require the combination with QCD corrections \cite{Altarelli:1978id,Altarelli:1979ub,Anastasiou:2003yy, Bozzi:2010xn,Duhr:2021vwj} and mixed QCD-electroweak corrections \cite{Armadillo:2024ncf,Buccioni:2022kgy,Bonciani:2021zzf}, as well as the simulation of QCD and QED radiation with MC methods (for either Drell-Yan production or $e^+e^-$ collisions), which is beyond the scope of this work.

\begin{table}[tb]
    \centering
    \begin{tabular}{|r@{\;=\;}l|r@{\;=\;}l|r@{\;=\;}l|}
    \hline
        $m_Z$ & $91.1535$ GeV & $\alpha^{-1}$ & 137.03599976 & $m_t$ & $173.2$ GeV \\
        ($m_Z^{\rm exp}$ & $91.1876$ GeV) & $\Delta \alpha$ & 0.059 & $\hat{m}_b$ & 4.18 GeV \\
        $m_W$ & $80.358$ GeV & $G_\mu$ & $1.1663787\times 10^{-5} \text{ GeV}^{-2}$ & $\hat{m}_c$ & 1.27 GeV \\
        ($m_W^{\rm exp}$ & $80.385$ GeV) & $m_\tau$ & 1.777 GeV & $\Bar{m}_s$ & $342\pm 123$ MeV \\
        $m_H$ & $125.1$ GeV & $m_\mu$ & 105 MeV & $\Bar{m}_{u,d}$ & $246\pm 113$ MeV \\
        \multicolumn{2}{|c|}{} & $m_e$ & 0.511 MeV & \multicolumn{2}{|c|}{} \\
        \hline
    \end{tabular}
    \caption{Input parameters for numerical evaluation. A hat ($\hat{m}$) indicates an $\overline{\text{MS}}$ mass, whereas a bar ($\Bar{m}$) denotes a threshold mass from Ref.~\cite{Erler:2017knj}.}
    \label{tab:input}
\end{table}

The chosen input parameter values are shown in Tab.~\ref{tab:input}. Light-fermion masses $m_{f \neq t}$ are neglected everywhere, except within the photon sub-loop self-energies insides of box diagrams, which were discussed at the end of the previous section. For the masses of the $u/d/s$ quarks, we use the so-called threshold masses, $\Bar{m}_{u,d,s}$, obtained from a renormalization-group analysis of the running electroweak couplings \cite{Erler:2017knj}. To be conservative, the associated quark mass uncertainties quoted in that reference were inflated by an additional factor of about 2. As mentioned previously, it is expected that the dependence on the light fermion masses drops out once the virtual NNLO corrections are combined with contributions with an additional light-fermion pair in the final-state, but we nevertheless show their impact in the plots for illustration.

\begin{figure}[tb]
\centering
   	\begin{tabular}{@{}cc@{}}   \includegraphics[height=0.25\textwidth]{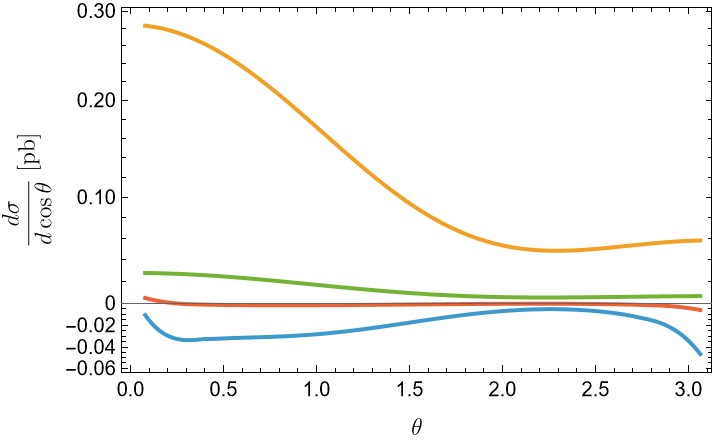} &  \includegraphics[height=0.25\textwidth]{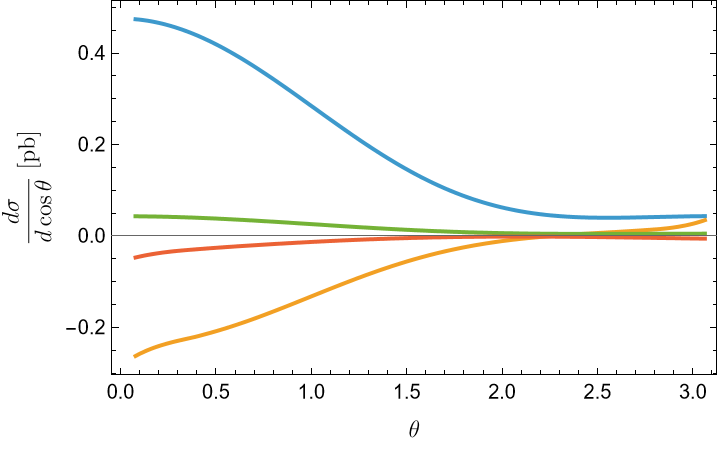} \\[-2ex] (a) $e^+e^-\to\mu^+\mu^-$ & (b) $e^+e^-\to u\bar{u}$ \\[1ex] \multicolumn{2}{c}{\includegraphics[height=0.25\textwidth]{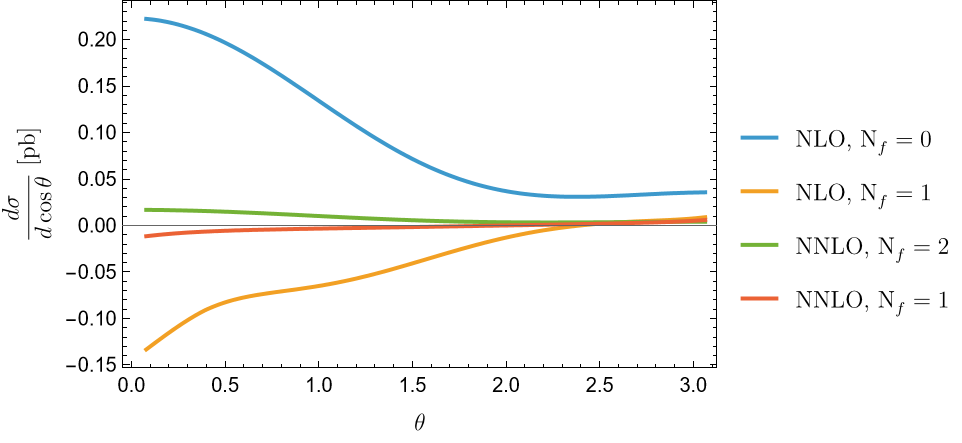}} \\[-2ex]
    \multicolumn{2}{c}{(c) $e^+e^-\to d\bar{d}$}
    \end{tabular}
    \vspace{-1ex}
  	\caption{Electroweak corrections to the differential unpolarized cross-section in the $\alpha(0)$ scheme at $\sqrt{s}=240$~GeV at NLO and NNLO. The width of the red curves reflects the light quark mass uncertainty.}
	\label{fig:diffxsec}
\end{figure}

\begin{figure}[p]
\centering
   	\begin{tabular}{@{}cc@{}}   \includegraphics[height=0.25\textwidth]{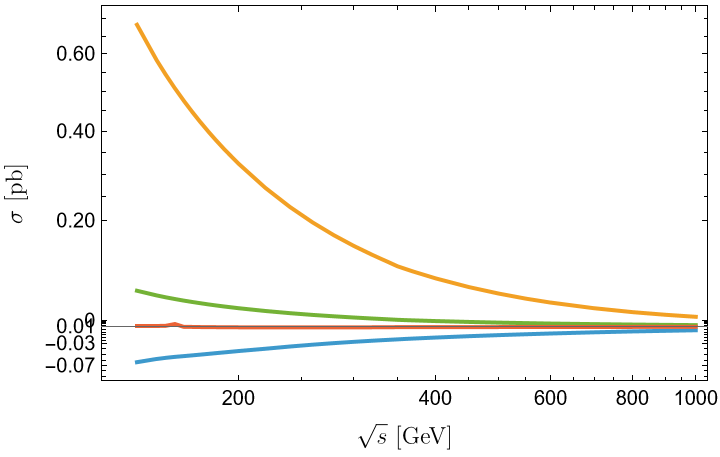} &  \includegraphics[height=0.25\textwidth]{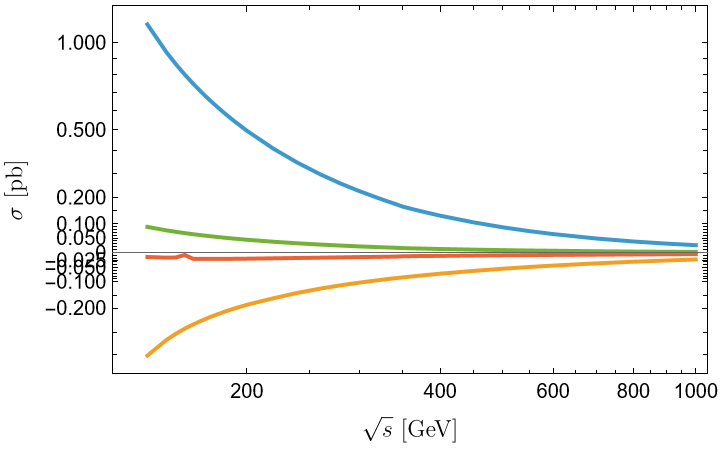} \\[-2ex] (a) $e^+e^-\to\mu^+\mu^-$ & (b) $e^+e^-\to u\bar{u}$ \\[1ex] \multicolumn{2}{c}{\includegraphics[height=0.25\textwidth]{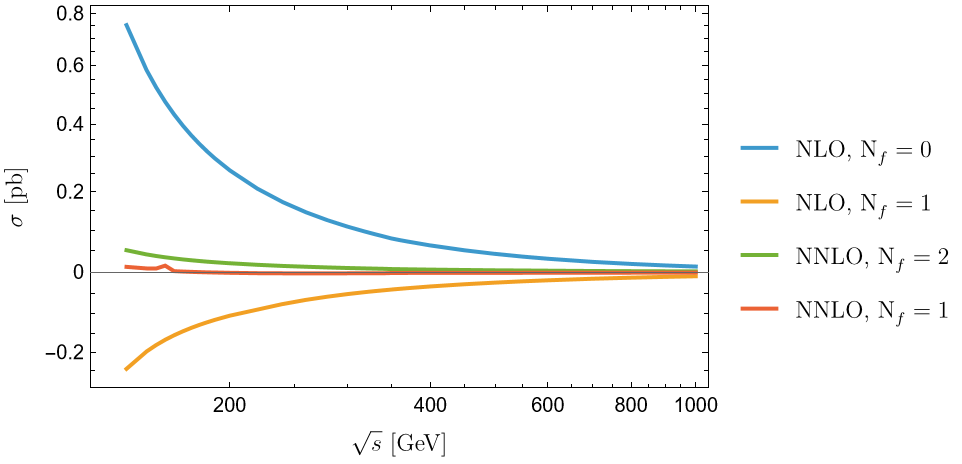}} \\[-2ex]
    \multicolumn{2}{c}{(c) $e^+e^-\to d\bar{d}$}
    \end{tabular}
    \vspace{-.5ex}
  	\caption{Electroweak corrections to the total unpolarized cross-section at NLO and NNLO in the $\alpha(0)$ scheme. The width of the red curves reflects the light quark mass uncertainty.}
	\label{fig:totxsec}
\end{figure}

In Figure~\ref{fig:diffxsec} we plot the unpolarized differential cross-section at NLO and NNLO for the three different final states ($\mu^+\mu^-$, $u\bar{u}$, $d\bar{d}$). Here and in the following, we will use the notation $N_f=n$ for diagram contributions with $n$ closed fermion loops. One can see that, depending on the kinematics and the initial- and final-state fermions, there are cancellations at NNLO between the diagrams with $N_f=1$ and $N_f=2$. 
In each of these plots, the uncertainty in the light quark masses is reflected in a band around the $N_f=1$ NNLO results.
Figure~\ref{fig:totxsec} shows total cross-section results for a range of center-of-mass energies. Pending the same factors as before, there is a cancellation between the $N_f=1$ and $N_f=2$ NNLO contributions, and the $N_f=1$ contributions show some features near the $WW$, $ZZ$ and $t\bar{t}$ thresholds. The electroweak NNLO corrections are typically at the level of $\lesssim 1\%$ of the LO cross-section, but their relative contribution tends to be enhanced at higher center-of-mass energies above a few hundred GeV.

\begin{figure}[p]
\centering
   	\begin{tabular}{@{}ccc@{}}   \includegraphics[height=0.25\textwidth]{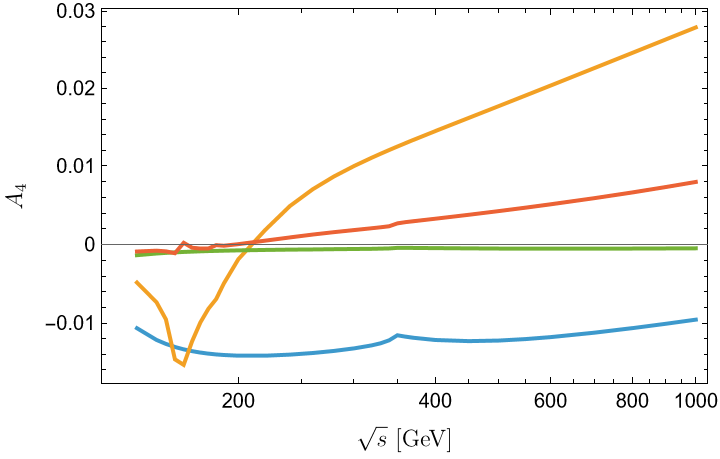} &  \includegraphics[height=0.25\textwidth]{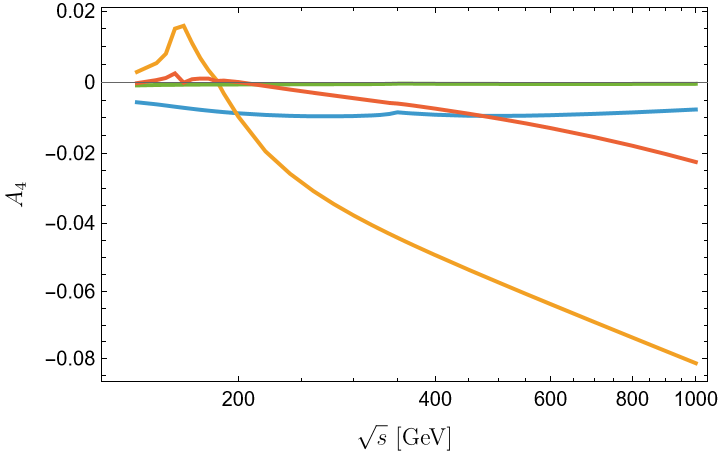} \\[-2ex] (a) $e^+e^-\to\mu^+\mu^-$ & (b) $e^+e^-\to u\bar{u}$ \\[1ex] \multicolumn{2}{c}{\includegraphics[height=0.25\textwidth]{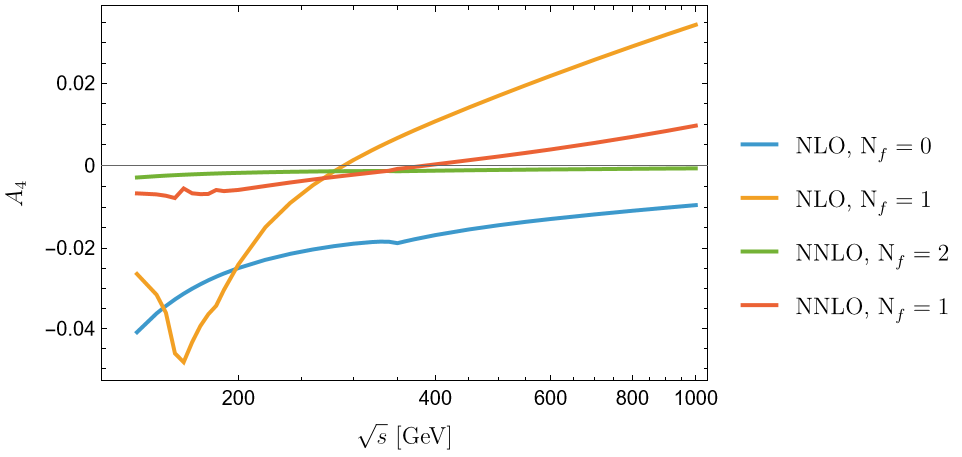}} \\[-2ex]
    \multicolumn{2}{c}{(c) $e^+e^-\to d\bar{d}$}
    \end{tabular}
    \vspace{-.5ex}
  	\caption{Similar to Fig.~\ref{fig:totxsec}, but for the $A_4$ parameter.}
	\label{fig:A4}
\end{figure}

Finally, Figure~\ref{fig:A4} plots the values of the $A_4$ parameter as defined as follows~\cite{Chaichian:1981va}:
\begin{align}
    A_4 &= \frac{8}{3}\,\frac{\sigma_{\rm F}-\sigma_{\rm B}}{\sigma_{\rm F}+\sigma_{\rm B}},
    \qquad
    \sigma_{\rm F} = \int_0^1 d\cos\theta\,\frac{d\sigma}{d\cos\theta}, \quad \sigma_{\rm B} = \int_{-1}^0 d\cos\theta\,\frac{d\sigma}{d\cos\theta}.
\intertext{Expanding this in perturbative orders yields}
    A_{4,\rm LO} &= \frac{8}{3} \;\frac{\sigma_{\rm F,LO}-\sigma_{\rm B,LO}}{\sigma_{\rm F,LO}+\sigma_{\rm B,LO}}, \\
    A_{4,\rm NLO} &= \frac{16(\sigma_{\rm B,LO}\sigma_{\rm F,NLO} - \sigma_{\rm F,LO}\sigma_{\rm B,NLO})}{3(\sigma_{\rm F,LO}+\sigma_{\rm B,LO})^2}, \label{eq:A4nlo} \\
    A_{4,\rm NNLO} &= \frac{16}{3} \biggl[ \frac{(\sigma_{\rm B,LO}\sigma_{\rm F,NNLO} - \sigma_{\rm F,LO}\sigma_{\rm B,NNLO})}{(\sigma_{\rm F,LO}+\sigma_{\rm B,LO})^2} - \frac{(\sigma_{\rm F,NLO}+\sigma_{\rm B,NLO})(\sigma_{\rm B,LO}\sigma_{\rm F,NLO} - \sigma_{\rm F,LO}\sigma_{\rm B,NLO})}{(\sigma_{\rm F,LO}+\sigma_{\rm B,LO})^3} \biggr]. \label{eq:A4nnlo}
\end{align}
The curves shown in the figures correspond to Eqs.~\eqref{eq:A4nlo} and \eqref{eq:A4nnlo}. It is interesting to note that for $A_4$ the NNLO corrections with $N_f=1$ are significantly larger than those with $N_f=2$, in contrast to the total cross-section.

Table.~\ref{tab:scheme} shows cross-section results at different perturbative orders for the different renormalization schemes for the electroweak coupling described in section~\ref{sec:calc}. A few representative values for the center-of-mass energy are chosen. ``NLO'' refers to the full NLO corrections, \emph{i.e.} the sum of $N_1=0$ and $N_f=1$ contributions, whereas ``NNLO'' denotes the sum of $N_f=2$ and $N_f=1$ contributions.
As can be seen from the table, the size of the higher-order corrections varies between the $\alpha(0)$, $\alpha(m_Z)$, and $G_\mu$ schemes. Large corrections in the former are due to numerically large terms proportional to powers of $\Delta\alpha$. More specifically, the size of the corrections is particularly small in the $\alpha(m_Z)$ scheme for $e^+e^-\to\mu^+\mu^-$, which can be understood from the fact that this process is dominated by photon exchange at tree-level. On the other hand, for $e^+e^- \to u\bar{u}/d\bar{d}$, the $G_\mu$ scheme is more effective at absorbing some of the higher-order corrections, since these processes have a significant Z-exchange component at tree-level.

\begin{table}[tb]
\small\centering
    \begin{tabular}{|l|r||c|c|c||c|c|c||c|c|c|}
    \hline
    \multicolumn{2}{|r||}{} & \multicolumn{3}{c||}{$e^+e^- \to \mu^+\mu^-$} & \multicolumn{3}{c||}{$e^+e^- \to u\bar{u}$} & \multicolumn{3}{c|}{$e^+e^- \to d\bar{d}$} \\
    \hline
    \multicolumn{2}{|r||}{Scheme} 
    & $\alpha(0)$ & $\alpha(m_Z)$ & $G_\mu$
    & $\alpha(0)$ & $\alpha(m_Z)$ & $G_\mu$
    & $\alpha(0)$ & $\alpha(m_Z)$ & $G_\mu$ \\
    \hline\hline
    $\sqrt{s}=$ & LO & 5.206 & 5.880 & 5.581 & 10.00 & 11.30 & 10.72 & 7.356 & 8.307 & 7.885 \\
    150~GeV & +NLO & 5.731 & 5.773 & 5.821 & 10.61 & 10.61 & 10.74 & 7.741 & 7.727 & 7.828 \\
     & +NNLO & 5.787 & 5.794 & 5.821 & 10.67 & 10.68 & 10.72 & 7.790 & 7.801 & 7.833 \\
    \hline
     & LO & 1.797 & 2.030 & 1.927 & 3.071 & 3.468 & 3.292 & 1.816 & 2.051 & 1.947 \\
    240~GeV & +NLO & 1.990 & 2.007 & 2.022 & 3.276 & 3.279 & 3.317 & 1.913 & 1.910 & 1.935 \\
     & +NNLO & 2.011 & 2.014 & 2.023 & 3.291 & 3.293 & 3.305 & 1.924 & 1.927 & 1.934 \\
    \hline
     & LO & 0.099 & 0.112 & 0.107 & 0.162 & 0.183 & 0.174 & 0.086 & 0.097 & 0.092 \\
    1~TeV & +NLO & 0.109 & 0.110 & 0.111 & 0.166 & 0.164 & 0.167 & 0.089 & 0.088 & 0.090 \\
     & +NNLO & 0.110 & 0.110 & 0.111 & 0.164 & 0.164 & 0.164 & 0.089 & 0.092 & 0.089 \\
    \hline
    \end{tabular}
    \caption{Partonic cross-sections in pb for different final states and in different renormalization schemes for the electroweak coupling.}
    \label{tab:scheme}
\end{table}

As mentioned above, the effective quark masses in Tab.~\ref{tab:input} are approximate parametrizations of non-perturbative hadronic effects in the photon sub-loop self-energy, and they have large uncertainties for the light quark masses, $\Bar{m}_{u,d,s}$. The impact of these quark mass uncertainties on our results is shown in Tab.~\ref{tab:errors}. They are generally below the per-mille level and thus negligible for practical purposes.

\begin{table}[tb]
\small\centering
    \begin{tabular}{|c|l|c|c|c|}
    \hline
    $\sqrt{s}$ & & $e^+e^- \to \mu^+\mu^-$ & $e^+e^- \to u\bar{u}$ & $e^+e^- \to d\bar{d}$ \\
    \hline\hline
& $\sigma_{\rm NNLO,ferm}$ [fb] & 5821.5 & 10724.7 & 7832.9 \\
     & $\Bar{m}_{u,d,s}$ uncertainty [fb] & 0.05 & 0.15 & 0.02 \\
    150~GeV & $\delta_{\rm perturb.,EW}$ from \eqref{eq:err23} [fb] & $7.1$ & $36.8$ & $10.3$ \\
     & scheme dep. [fb] & 27.7 & 42.7 & 31.7 \\
     & max.\ of prev.\ two rows [fb] & 27.7 (0.48\%) & 42.7 (0.40\%) & 31.7 (0.41\%) \\
    \hline
    & $\sigma_{\rm NNLO,ferm}$ [fb] & 2023.1 & 3305.1 & 1933.8 \\
     & $\Bar{m}_{u,d,s}$ uncertainty [fb] & 0.02 & 0.06 & 0.008 \\
    240~GeV & $\delta_{\rm perturb.,EW}$ from \eqref{eq:err23} [fb] & 3.4 & $21.4$ & $6.1$ \\
     & scheme dep. [fb] & 9.3 & 12.4 & 7.3 \\
     & max.\ of prev.\ two rows [fb] & 9.3 (0.46\%) & 21.4 (0.65\%) & 7.3 (0.38\%) \\
    \hline
    & $\sigma_{\rm NNLO,ferm}$ [fb] & 110.61 & 164.43 & 89.38 \\
     & $\Bar{m}_{u,d,s}$ uncertainty [fb] & 0.002 & 0.006 & 0.001 \\
    1~TeV & $\delta_{\rm perturb.,EW}$ from \eqref{eq:err23} [fb] & 0.78 & 5.6 & 1.4 \\
     & scheme dep. [fb] & 0.36 & 0.17 & 0.16 \\
     & max.\ of prev.\ two rows [fb] & 0.78 (0.7\%) & 5.6 (3.4\%) & 1.4 (1.6\%) \\
    \hline
    \end{tabular}
    \caption{Evaluation of uncertainties from light quark masses and missing higher orders, using the $G_\mu$ scheme as reference. See text for more details.}
    \label{tab:errors}
\end{table}

Finally, we comment on the uncertainty from missing higher-order electroweak corrections. We use two different methods for evaluating the missing NNLO $N_f=0$ contributions. Method (a) only looks at $N_f=0$ corrections are different orders and assumes that the ratio of NNLO to NLO is similar to the ratio of NLO to LO, resulting in the estimate
\begin{align}
    \delta_a \sigma_{{\rm NNLO},N_f=0} \sim \bigl(\delta \sigma_{{\rm NLO},N_f=0} \bigr)^2 / \sigma_{\rm LO} . \label{eq:err2a}
\end{align}
Method (b) assumes that the ratio of $N_f=1$ and $N_f=0$ contributions is similar at NLO and NNLO, leading to the estimate
\begin{align}
    \delta_b \sigma_{{\rm NNLO},N_f=0} \sim \Biggl|\delta\sigma_{{\rm NNLO},N_f=1} \frac{\delta \sigma_{{\rm NLO},N_f=0}}{\delta \sigma_{{\rm NLO},N_f=1}}\Biggr| . \label{eq:err2b}
\end{align}
For third-order (N3LO) electroweak correction, we also use a scaling argument for the perturbative series:
\begin{align}
    &\delta\sigma_{\rm N3LO} \sim \bigl( \delta'\sigma_{\rm NNLO} \bigr)^2 / \delta'\sigma_{\rm NLO} \label{eq:err3}
\intertext{where}
    &\delta'\sigma_{\rm NNLO} \equiv \sqrt{\sigma_{{\rm NNLO},N_f=2}^2 + \sigma_{{\rm NNLO},N_f=1}^2}\,,
    \quad
    &\delta'\sigma_{\rm NLO} \equiv \sqrt{\sigma_{{\rm NLO},N_f=1}^2 + 
    \sigma_{{\rm NLO},N_f=0}^2}\,.
    \label{eq:err3a}
\end{align}
The construction in eq.~\eqref{eq:err3a} is used to circumvent the impact of accidental cancellations between different $N_f$ contributions.
To arrive at an estimate of the overall perturbative uncertainty, we take the maximum of \eqref{eq:err2a} and \eqref{eq:err2b} and combine it in quadrature with \eqref{eq:err3},
\begin{align}
    \delta\sigma_{\rm perturb.,EW} = \sqrt{\bigl(\max\bigl\{\delta_a \sigma_{{\rm NNLO},N_f=0},\delta_b \sigma_{{\rm NNLO},N_f=0}\bigr\}\bigr)^2 + \delta\sigma_{\rm N3LO}^2}\,. \label{eq:err23}
\end{align}

Alternatively, one can also use the scheme dependence between the $\alpha(m_Z)$ and $G_\mu$ schemes as a proxy for the missing higher order uncertainty. Both estimates are reported in Tab.~\ref{tab:errors}, and we conservative take the maximum of \eqref{eq:err23} and the scheme dependence as the intrinsic theoretical uncertainty from higher-order electroweak corrections.

As evident from Tab.~\ref{tab:errors}, the electroweak perturbative uncertainties are generally at the per-cent level or below for $\sqrt{s} \sim \text{few } 100$~GeV, but they become larger at TeV-scale center-of-mass energies. This behavior is expected due to the impact of electroweak Sudakov logarithms $\sim \alpha^n\ln^{2n}(s/m_W^2)$~\cite{Fadin:1999bq,Denner:2000jv,Denner:2001gw,Kuhn:2001hz,Beenakker:2001kf,Jantzen:2005az,Denner:2006jr}.


\section{Summary}
\label{sec:summ}
This article reports on the evaluation of complete fermionic electroweak two-loop corrections to the Drell-Yan and related fermion-pair production processes. The calculation has been carried out using a dispersion relation technique to evaluate the contributions of vertex and box diagrams with fermion self-energies or triangle sub-loops. UV divergences were handled by subtracting analytically integrable terms to create UV-finite and numerically stable expressions. IR divergences were removed using universal eikonal QED factorization. Even after this factorization, we observed a remaining sensitivity of the virtual two-loop corrections on non-perturbative hadronic contributions, which we estimated by using effective light-quark masses and found to be numerically negligible.

We evaluated the fermionic electroweak corrections to $e^+e^- \to \mu^+\mu^-/u\bar{u}/d\bar{d}$ and compared them to the previous orders. We found that, for a representative center-of-mass energy of $\sqrt{s}=240$~GeV in the $\alpha(0)$ scheme the fermionic NNLO corrections are about 1\% or less of the NLO results for the various final states. Depending on the specific process and collision energy, the corrections can be smaller in the $\alpha(m_Z)$ and $G_\mu$ schemes but are generally of the same order. We also estimated the remaining higher-order electroweak corrections, finding that they would be on the per-cent order for center-of-mass energies up 1~TeV.

To fully assess the impact of the electroweak NNLO corrections on Drell-Yan production at the LHC or other hadron collider, one needs to combine them with the available QCD and mixed QCD-electroweak corrections, parton distribution functions, and MC generators for the simulation of QCD and QED radiation. We leave this for future work.


\section*{Acknowledgments}

The authors would like to thank T.~Armadillo, S.~Devoto and A.~Vicini for useful discussions and comparison of some partial results.
This work has been supported in part by the U.S.~National Science Foundation Grant No.\ PHY-2412696.


\bibliographystyle{JHEP}{}
\bibliography{ew2f}

\end{document}